\documentclass[twocolumn]{article}

 \usepackage[pdftex]{graphicx}

\usepackage[cmex10]{amsmath}

\usepackage{array}

\usepackage{fixltx2e}
\begin{document}

\title{An Information-Theoretical Approach to Image Resolution Applied to Neutron Imaging Detectors based upon Individual Discriminator Signals.}

\author{Jean-Fran\c{c}ois Clergeau, Matthieu Ferraton\thanks{Clergeau, Ferraton, Gu\'{e}rard, Khaplanov, Piscitelli, Platz, Rigal and Van Esch are with the Institut Laue Langevin in Grenoble, France.}, \\
Bruno Gu\'{e}rard, Anton Khaplanov, \\
Francesco Piscitelli, Martin Platz, \\
Jean-Marie Rigal, Patrick Van Esch \thanks{Corresponding author vanesch@ill.fr.} ,\\ 
Thibault Daull\'{e}
\thanks{Daull\'{e} is with PHELMA Grenoble - INP.}
}

\maketitle

\begin{abstract}
1D or 2D neutron imaging detectors with individual wire or strip readout using discriminators have the advantage of being able to treat several neutron impacts partially overlapping in time, hence reducing global dead time.  A single neutron impact usually gives rise to several discriminator signals.  In this paper, we introduce an information-theoretical definition of image resolution.  Two point-like spots of neutron impacts with a given distance between them act as a source of information (each neutron hit belongs to one spot or the other), and the detector plus signal treatment is regarded as an imperfect communication channel that transmits this information.  The maximal mutual information obtained from this channel as a function of the distance between the spots allows to define a calibration-independent measure of resolution.  We then apply this measure to quantify the power of resolution of different algorithms treating these individual discriminator signals which can be implemented in firmware.    The method is then applied to different detectors existing at the ILL.  Center-of-gravity methods usually improve the resolution over best-wire algorithms which are the standard way of treating these signals.
\end{abstract}

\section{Introduction}
A neutron detector that operates in counting mode has several important quality parameters.  Those parameters can be divided in two classes.  The first class is related to its ability to detect genuine neutrons and reject all other kinds of signals or noise.  They are the quantum efficiency, the background counting rate, the highest counting rates and associated dead time and gamma sensitivity.  Other parameters play a role in the image quality, but in this paper we will turn to the original meaning of one such parameter: resolution, the ability to discriminate two nearby points in the image.

The way this is usually treated is by making the hypothesis that the image of a point source is the point spread function (PSF), and the resolution is then nothing else but a study of the width of this PSF, which can be done in the real domain or in the frequency domain.  However, the use of a PSF makes the underlying hypothesis that the imaging system is translation-invariant: in other words, that the image of a point has the same form (the PSF) no matter where it is located within the image.  This can be a very good approximation for 'continuous' imaging systems (such as chemical film imaging), or when the digitizing pixels are much smaller in size than the width of the PSF.

Neutron detectors usually have a large granularity in their construction, and the resolution is often of the same size as the detector granularity.  In such a case, the hypothesis of translation invariance on the level of the detector granularity or below is not valid anymore, and the concept of PSF becomes delicate to handle.  The image of a point source (and its width) will depend on the exact location of the point with respect to the grid defining the physical detector granularity.

One can consider the width measure of the image of a point source as a function of its location as a measure of 'local resolution'.  This value is usually periodic with the same period as the detector granularity.  However on further analysis, it should be evident that this is problematic.  Indeed, resolution is defined as the ability to distinguish two point sources as a function of their distance.  It doesn't make much sense to define locally a resolution that varies faster from point to point than its own value!  If the 'local resolution' has gone through a whole period of values over the size of the 'resolution', it is absolutely not clear which of these values indicates the distance over which two point images can be resolved.

In order to define image resolution, one needs to consider \emph{two} images, of two point sources, and determine how different they are.  The image of a single point can only give information about resolution if extra hypotheses are made, such as translation-invariance.  If thas hypothesis doesn't hold, then the resolution will differ locally from point, and needs to be established by comparing pairs of images taken at different points.  If one wants to specify a global resolution parameter, it seems natural to take the worst case as this value is guaranteed to hold for all pairs of points, independent of their exact location with respect to the granularity grid of the detector.

\section{Information-Theoretical Definition of Image Resolution}

\subsection{Resolution and Point Spread Function}

Under the hypothesis of translation-invariance, continuity and linearity, the image function $P$ is obtained from the source function $S$ by convolution with the PSF \cite{wolf:psf} :
\begin{equation}
\label{eq_convolutionPSF}
P(x,y) = \int_{u}\int_{v} S(u,v) PSF(x-u,y-v) du dv
\end{equation}
In the frequency domain, this can be written:
\begin{equation}
\label{eq_MTF}
P_f(\omega_x,\omega_y) = S_f(\omega_x,\omega_y) MTF(\omega_x,\omega_y)
\end{equation}
The Fourier Transform of the PSF is called the modulation transfer function (MTF).  In the ideal case, $PSF(x,y)$ reduces to $\delta^2(x,y)$ and we obtain that $P(x,y) = S(x,y)$: the image is a faithful representation of the source.  The MTF then becomes unity for all values of $\omega_x$ and $\omega_y$. 

In the real world, all imaging equipment has, for technical of fundamental physical reasons, a finite bandwidth in the frequency domain, and thus also a finite PSF width.  Any point source $\delta^2(x-x_0,y-y_0)$ gives the PSF as an image, centered on the point $(x_0,y_0)$.

One defines resolution usually as one or other measure of the width of the PSF, or a measure of the bandwidth of the MTF.  Both are related.  A popular measure is the Full Width Half Maximum (FWHM) criterion of the PSF, as described in \cite{fedstan:1037C} and \cite{quant:zaidi}.   Making the hypothesis that the PSF has a Gaussian shape, the FWHM value equals 2.35 times the standard deviation of the PSF.  By extrapolation, one defines the pseudo-FWHM value of any PSF as 2.35 times the standard deviation of that PSF.

All these concepts, which make perfect sense under the stated hypotheses of linearity, continuity and translation-invariance, start losing their meaning when we have a granularity of the size of the resolution.

\subsection{Resolution Revisited}

In a particle-counting detector, the images discussed in the previous sub-section have to be interpreted as probability distributions of detection for each individual particle.  For neutron detectors, any conceivable resolution and granularity is many orders of magnitude larger than the quantum-mechanical wavelength of the neutron at hand (thermal neutrons having a wavelength of 1.8 \AA), so we don't have to take into account any quantum interference effect.  If we consider two point sources, we can represent this as two random streams of neutrons: all the neutrons in one stream will impact on position 1, and all the neutrons in the second stream will impact on position 2.  The randomness resides in the fact that we don't know whether the next neutron belongs to stream 1 or to stream 2.  After a long acquisition time, all the neutrons in stream 1 will give an image corresponding to the image of a point source at point 1 ; all the neutrons in stream 2 will give an image corresponding to the image of a point source at point 2.

We want to define resolution as our ability to determine, from the measured impact position of a neutron, whether it belonged to stream 1 or stream 2, as a function of the distance between points 1 and 2.

If the two images (of a point source at point 1, and a point source at point 2 respectively) are entirely distinct, then it will not be difficult to set up a criterion upon impact of a neutron, to determine whether it belonged to stream 1 or stream 2: it is sufficient to choose a separation in the image somewhere between the two distributions: if the impact is to the left of our chosen separator, we say that it is a neutron of stream 1, and if it is to the right, we say that it belonged to stream 2.  In as much as the distributions are strictly distinct, the probability to make a mistake is 0.  However, when the two distributions overlap partly, it is harder to define a criterion, and in any case we will make mistakes: we will assign certain neutrons to stream 2 whereas they actually belonged to stream 1 and vice versa.  We would like to optimize our separation criterion.

The problem at hand is in fact identical to the problem of a lossy binary communication channel.  A binary communication channel consists in the transmission of a random bitstream, where on the average there are as many 0 as 1, and the receiver has to reconstruct the incoming bit stream as faithfully as possible, but the channel is noisy, and some emitted bit 0 will be detected as a bit 1 and vice versa.

The analogy is clear: a neutron in stream 1 corresponds to the emission of a bit 0, and a neutron in stream 2 corresponds to the emission of a bit 1.  The detector is the receiver of our channel, and tries to find out whether we had a bit 0 or a bit 1 by applying a separation criterion.

A lossy communication channel is determined by 2 independent probabilities: the probability that an emitted 0 is correctly detected as 0 ; and the probability that an emitted 1 is correctly detected as a 1.  We can represent this in a table:
\begin{center}
\begin{tabular}{||c|c|c|c||}
\hline
sent & received 0 & received 1 & sum \\
\hline
0 & $p(0|0)$ & $p(1|0)$ & $= 1$ \\
1 & $p(0|1)$ & $p(1|1)$ & $= 1$ \\
\hline
\end{tabular}
\end{center}
Note that this table contains conditional probabilities.  The joint probability is half these values because we make the assumption that the probability to send a 0 is $1/2$ and the probability to send a 1 is $1/2$.

A lossy binary channel has a mutual information as defined in \cite{mathcom:ShannonWeaver}:
\begin{equation}
\label{eq_defI}
I(X;Y) = \sum_{x,y} p(x,y) \log_2 \frac{p(x,y)}{p(x)p(y)}
\end{equation}
It gives us the amount of information we have about the random variable $Y$ (the emitted signal) by the knowledge of the random variable $X$ (the received message).  Note that the expression is symmetric in $X$ and $Y$.  Mutual information is often used in medical imaging to extract common information from different images \cite{quant:zaidi}.  We will use it in a slightly different context.

In the case of a binary signal with a priori probabilities $1/2$, this expression can be written as a function of the conditional probabilities only:
\begin{equation}
\label{eq_binaryI}
I(X ; Y) = \frac{1}{2} \sum_{y=0}^1 \sum_{x=0}^1 p(x|y) \log_2 \frac{p(x|y)}{\frac{1}{2}\left( p(x|0) + p(x|1)\right)}
\end{equation}
Transposed to our image resolution problem, after we have defined a separation criterion in the image, $I$ represents the information we have obtained from a single neutron impact about the point source from which it originated.

By resolution, we can then understand the distance between two point sources needed in order to reach a given threshold of information $I$ when using the best possible separation criterion.

\subsection{Method}

We consider a set of images of single point sources at different positions $x_i$ ; these images are represented as (discretized) normalized probability distributions $D_i(u)$ where we take $u$ to run over the bins from 0 to $N$.  We now consider all the couples $(i,j)$ of such images, which correspond to two point sources, one at position $x_i$ and another at position $x_j$.  We keep the index $i$ fixed, and let the index $j$ run.  For a given couple $(i,j)$, we consider all possible discrete separators at positions $s$.  For a given separator $s$, we consider an impact to the left of $s$ as coming from source $i$ and an impact to the right of $s$ to be coming from source $j$.  With the images $D_i$ and $D_j$, we can calculate the 4 conditional probabilities:
\begin{eqnarray}
p(0|0) & = \sum_{u=0}^{s-1} D_i(u) \label{eq_pstart} \\
p(0|1) & = \sum_{u=0}^{s-1} D_j(u) \\
p(1|0) & = \sum_{u=s}^N D_i(u) \\
p(1|1) & = \sum_{u=s}^N D_j(u) \label{eq_pend}
\end{eqnarray}
and from this table, we can calculate the mutual information which we will write $I(i,j,s)$.  If the separator is badly chosen, $I$ will be close to 0, but for the right value of $s$, $I$ will go through a maximum: that separator $s$ will be the best possible separator in order to maximize our information about from which point source $i$ or $j$ the neutron emerged.  This maximum is called $\dot{I}(i,j)$.  If we now consider $\dot{I}(i,j)$ as a function of the running $j$ with $i$ fixed, we normally get a monotonic rising function ; the more the two spots $x_i$ and $x_j$ are separated, the more information we can extract about the origin of the point source for each neutron impact.

When $\dot{I}(i,j)$ reaches a certain threshold $I_{res}$, we consider that the distance $|x_i - x_j|$ is equal to the resolution: we define resolution of an imaging system as the distance necessary between two point sources so that each impact from these sources results in more than $I_{res}$ bits of information about its origin.

\subsection{Link with PSF}

\begin{figure}[!t]
\centering
\includegraphics[width=3.5in]{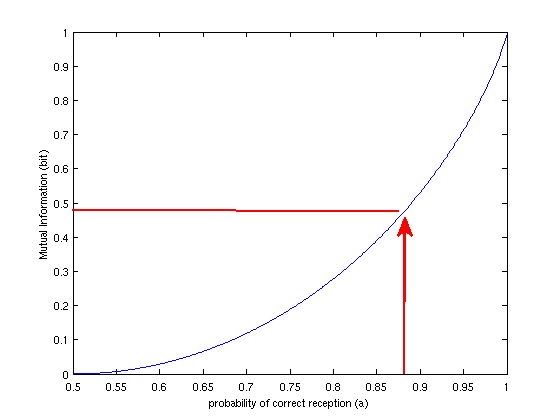}
\caption{Relationship between mutual information (in bits) and probability of correct reception in the symmetric, translation-invariant case.}
\label{fig_Iversusa}
\end{figure}

It is important to link our information-theoretical criterion with the more standard definitions of resolution based upon the width of the PSF.  Under the assumption of translation-invariance, all the $D_i(u)$ have the same shape, namely that of the PSF, centered at $x_i$.  If moreover we assume that the PSF is symmetrical, it is easy to show that the optimal separator is half-way between the two points $x_i$ and $x_j$.  With no loss in generality, we can put $x_i$ to 0 and $x_j = d$, where $d$ is the distance between the two point sources.  As such it is easy to show that, if we put $a$ equal to 
\begin{equation}
a = \int_{u = -\infty}^{d/2} PSF(u) du
\end{equation}
that
\begin{eqnarray}
p(0|0) & = & a \label{eq_pastart}\\
p(0|1) & = & 1 - a \\
p(1|0) & = & 1 - a \\
p(1|1) & = & a \label{eq_paend}
\end{eqnarray}
Here, $a$ is the probability to make the right reception of the message.  Note that it doesn't make sense to consider $a < 0.5$, because then we simply inverse sources 1 and 0.  When $a = 0.5$, the received message doesn't contain any information regarding the emitted signal.  It is easy to substitute the values in equations \ref{eq_pastart} to \ref{eq_paend} in equation \ref{eq_binaryI} to obtain:
\begin{equation}
\label{eq_Iasfunctionofa}
I(a) = a \log_2(a) + (1 - a) \log_2(1-a) + 1
\end{equation}
This relationship is shown in figure \ref{fig_Iversusa}.

Using the FWHM criterion, we find that $a = 0.88$ ; in other words, if the PSF is Gaussian, and the two point sources are separated by a distance given by the FWHM, we have 88\% chance to identify correctly the origin of each impact of a neutron.  Using equation \ref{eq_Iasfunctionofa}, this leads to a threshold value of mutual information equal to 0.47 bits.   We will use this criterion hence as the $I_{res}$ to use.

It should be pointed out also that the FWHM criterion (or for that matter any other 'classical' PSF width criterion) needs the PSF to be expressed in physical width units, and hence is dependent on a position calibration.  Under the assumption of translation-invariance, this calibration is independent on the position, that is, there is no 'linearity error'.   A real detector can have a non-linear calibration, especially when the position is discretized, and this non-linearity should ideally be taken into account in the conversion of the FWHM of the image (in discretized form of course) to a physical distance representing resolution.  In other words, the center of gravity of the discretized image of a point source doesn't evolve proportionally with the physical position of the point source. 
However, as pointed out earlier, the very fact that this non-linearity changes over the distance of the resolution itself, makes the whole concept delicate to use as such.  This is why we specify the FWHM in 'units of bins' without attempting to convert it to a physical distance other than the nominal bin size. 

The advantage of the mutual information criterion is that it is totally independent of any kind of calibration: we compare two images seen as probability distributions and we don't need to know the calibration of the horizontal axis.  The mutual information depends just on the degree of separability of the two images.  

\section{Neutron Detectors with Individual Discriminator Signals}

\subsection{The Algorithms}

Position sensitive neutron detectors can consist of many individual electrodes which can each have their own readout circuit.  In the simplest of cases, this readout electronics consists of the standard charge amplifier, shaper, and discriminator chain.  The logical discriminator signal conveys two pieces of information for each pulse: the time of the crossing by the shaped pulse of the threshold value, and the Time over Threshold (ToT) of this signal.  This time over threshold gives some information concerning the amount of deposited charge on the electrode.

The thermal neutron detectors we consider here are all of the type proportional gas detectors, with He-3 as a converter gas.  We consider MWPC and MSGC detectors in 1 and 2 dimensions. 

A single neutron impact can trigger signals on several electrodes, and the precise timing information and time-over-threshold (ToT) information still present in the discriminator signals can be used to determine a more accurate impact position.  This treatment is done by an online algorithm ; in our case this algorithm is programmed in one or several FPGA units processing the logical discriminator  signals. The output of this algorithm assigns a neutron impact to a specific, discretized bin.

All else equal, the choice of the algorithm will have an influence on the image quality: the more precise the impact position reconstruction by the algorithm, the better the image quality will be.  The algorithms we will consider are meant to reconstruct as well as possible the center of gravity of the primary charge cloud in the detector.  There is no attempt to reconstruct the asymmetrical tracks of the proton and the triton in order to compensate the distance between the impact position and the charge cloud center, so this difference (linked to the gas composition and pressure) will constitute a lower limit of the achievable resolution in any case.
\begin{description}
\item[FA]The first active algorithm is used often because of its easy implementation with elementary logic circuits: we pick the electrode that first crosses the threshold value, blocking out the neighboring electrode signals.  The idea is that the largest charge (in the center of the cloud) will give rise to the highest shaped pulse which will hence cross the fixed threshold value somewhat earlier than another, smaller charge, deposited at the same time.  The problem with this algorithm is that the charges are physically not deposited at the same time on the electrodes, giving an advantage to early deposits which do not necessarily correspond to the centre of the charge cloud.  Another problem is that small differences in propagation delay between different electronic signals can systematically give the advantage to some channels over their neighbours.
\item[MaxToT] 
In order to remedy against the problems of the first active algorithm, the longest ToT is a more robust criterion to find the electrode with the largest amount of charge on it, making this algorithm more robust against small variations in the electronics parameters of neighbouring channels.  The disadvantage remains that the largest signal doesn't necessarily correspond to the center of the charge cloud.
\item[CoG ToT] 
In this algorithm, the electrode positions are weighted with their ToT values to calculate a Center of Gravity.  This center of gravity can be digitized in smaller bins than the individual electrode numbers.  The advantage of this algorithm is that we really try to reconstruct the center of the charge cloud.  The disadvantage is the much heavier processing online, and also the fact that the ToT is not proportional to the charge, introducing a non-linearity error in the calculation of the center of gravity.  We have chosen to digitize this calculated center of gravity onto twice as many bins as there are physical channels.
\item[LR FA] Left-right first active algorithm.
With this algorithm, we consider the first active signal to select the main electrode, but we associate two bins to each electrode, picking the side of the second fastest signal.  The algorithm is lighter to implement, and corrects partly the sensitivity of the brute first active signal algorithm.  
\end{description}

The First Active and the Maximum ToT algorithms have a discretisation which is of course equal to the physical detector cell size.  The two other algorithms have twice as much bins (of half the physical cell size).  Each physical cell is divided in two (a 'left' and a 'right' half cell), and the algorithm specifies which of the two half-cells is elected.  In the case of indetermination, because only one physical cell is hit, one of the two half cells is selected randomly with a probability of 0.5. 

We will determine the influence of the choice of the algorithm on the resolution for two different types of detectors.  It should be obvious that in the case only one physical cell is hit, there is no difference between the 4 algorithms: the first two algorithms will select the cell at hand, and the last two algorithms will divide equally (by random selection each time) the hits over the two half-sized bins belonging to the physical cell at hand.  The algorithms can only bring in extra information when multiple physical cells are hit.

\subsection{The D4 Detector at the ILL}

The D4 detector consists of an array of 9 individual 1D MSGC detectors containing 15 bar of helium.  The MSGC plates have anode-cathode cells which repeat every 2.5 mm.  This instrument is more fully described in \cite{d4c:fischer}.

In order to have sufficient counting rate (and hence a good signal to background ratio in the image) with a narrow beam (less than 1 mm wide), we used a narrow rectangular beam (image) instead of a point image. We scanned 10 mm of detector area in steps of 0.2 mm, and we applied the 4 different algorithms in these scans.  We applied a simple constant background subtraction to each of the images.

\begin{figure}[!t]
\centering
\includegraphics[width=3.5in]{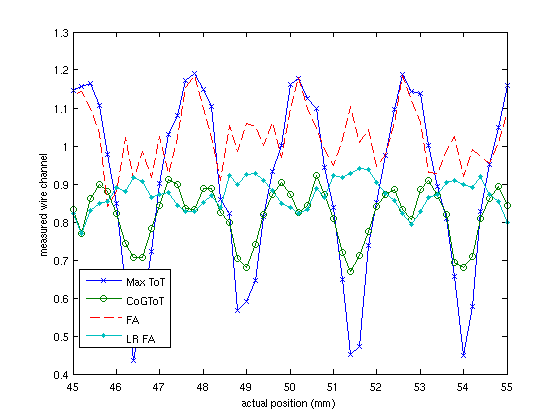}
\caption{The estimated pseudo FWHM of each point source image as a function of position on the D4 detector.}
\label{fig_pseudofwhmD4}
\end{figure}

\begin{figure}[!t]
\centering
\includegraphics[width=3.5in]{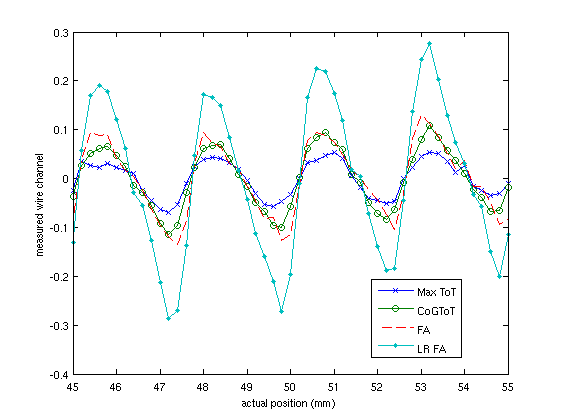}
\caption{The linearity error of the centers of gravity of the point source images as a function of the real position of the point source obtained after subtraction of the best linear fit for the D4 detector.}
\label{fig_calibnonlinD4}
\end{figure}

The standard analysis of the pseudo-FWHM as a function of position is shown in figure \ref{fig_pseudofwhmD4}.  The non-linearity of the calibration is of a similar order of magnitude as the differences between the pseudo-FWHM values for different algorithms, so it is difficult to establish a genuine value for the spatial resolution per algorithm.  Nevertheless we see that the two algorithms with two bins per physical channel have a sensibly lower maximum FWHM (namely about 0.9 physical channels) than the algorithms with only one bin (about 1.2 physical channels), which may indicate that we win about 25\% in worst case resolution.  Of these two, the center of gravity of ToT algorithm has the lowest non-linearity (which is of the same order as the non-linearity of the FA algorithm and only slightly worse than the Max ToT algorithm) which makes it the prefered algorithm concerning resolution when studied with standard methods.

\begin{figure}[!t]
\centering
\includegraphics[width=3.5in]{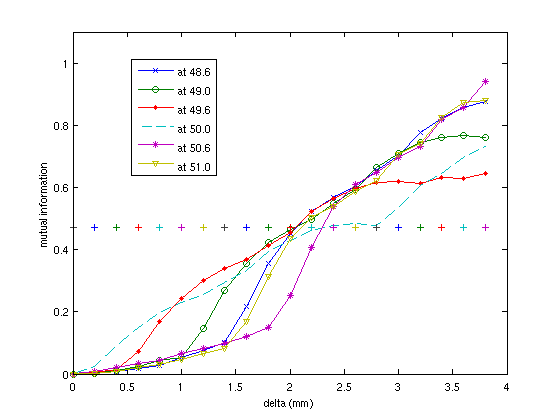}
\caption{The mutual information as a function of the distance between the point sources when the FA algorithm is applied to the D4 detector data.  The crosses represent the $I_{res}$ value of 0.47 corresponding to a pseudo-FWHM separation.}
\label{fig_infocurvest0D4}
\end{figure}

\begin{figure}[!t]
\centering
\includegraphics[width=3.5in]{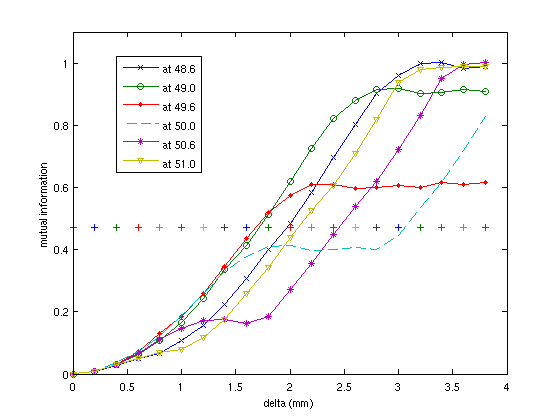}
\caption{The mutual information as a function of the distance between the point sources when the max ToT algorithm is applied to the D4 detector data.  The crosses represent the $I_{res}$ value of 0.47 corresponding to a pseudo-FWHM separation.}
\label{fig_infocurvesmaxToTD4}
\end{figure}

\begin{figure}[!t]
\centering
\includegraphics[width=3.5in]{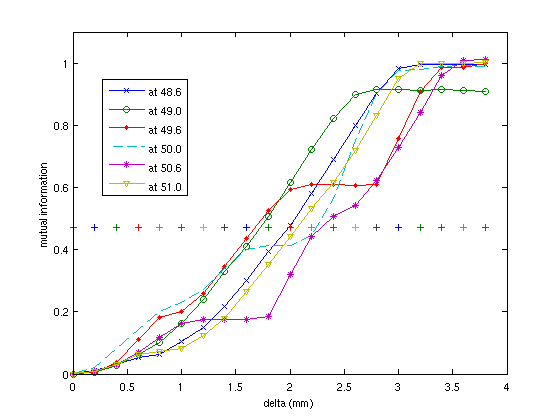}
\caption{The mutual information as a function of the distance between the point sources when the CoG ToT algorithm is applied to the D4 detector data.  The crosses represent the $I_{res}$ value of 0.47 corresponding to a pseudo-FWHM separation.}
\label{fig_infocurveslRCoGToTD4}
\end{figure}

\begin{figure}[!t]
\centering
\includegraphics[width=3.5in]{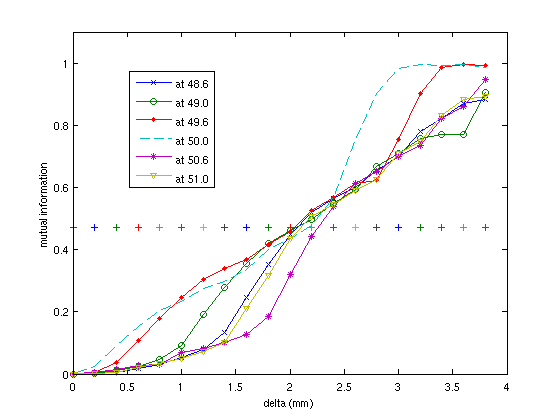}
\caption{The mutual information as a function of the distance between the point sources when the LR FA algorithm is applied to the D4 detector data.  The crosses represent the $I_{res}$ value of 0.47 corresponding to a pseudo-FWHM separation.}
\label{fig_infocurveslRt0D4}
\end{figure}

When calculating the mutual information as a function of source distance using a given algorithm, we find curves such as in figure \ref{fig_infocurvest0D4} for the FA algorithm, figure \ref{fig_infocurvesmaxToTD4} for the max ToT algorithm, figures \ref{fig_infocurveslRCoGToTD4} and \ref{fig_infocurveslRt0D4} respectively for the CoG ToT and the LR FA algorithms.  The different curves on the same plot correspond to different starting positions.  From these plots we can conclude that the FA algorithm obtains a resolution of 2.0 mm to 2.3 mm (but this could easily be 2.8 mm too), the max ToT algorithm obtains a resolution between 1.7 mm and 3 mm, the CoG ToT algorithm from 1.7 mm to 2.3 mm and the LR FA results in a resolution going from 2.0 mm to 2.3 mm.

The CoG ToT algorithm combines both the lowest lower bound and the lowest higher bound and is thus optimal concerning resolution as we defined it.  It can easily be checked that the mutual information curve of the CoG ToT algorithm is in almost all cases the highest curve as compared to the information curves from the other algorithms.

\subsection{The BIDIM-26 Detector at the ILL}

The Bidim-26 detector is a bi-dimensional position-sensitive MWPC detector with an active surface of about 26 cm $\times$ 26 cm.  It consists of a cathode wire plane with channel steps\footnote{The cathode plane actually consists of wires with a step of 1 mm, but they are grouped in physical channels by connecting them two by two.} of 2 mm, and an anode wire plane perpendicular to it, also with a step of 2 mm.  Both the anode and cathode wire planes are individually read out.  In our setup, the anode wires are vertical, and the cathode wires are horizontal.  The coincidence of an event on the anode wire plane and on the cathode wire plane allows to have 2-dimensional information.

We will examine the projections of the image along the anode wires and along the cathode wires individually by using a narrow vertical beam (exploring the 'anode' resolution) and a narrow horizontal beam (exploring the 'cathode' resolution) respectively.  The comparison can be interesting because the average multiplicity of hits (the average number of wires hit by a neutron event) is larger for the cathode plane than for the anode plane, while for the rest, all the physics in the detector is of course identical (it is the same physical charge that induces both signals on anode and cathode).  In the study of this detector, we didn't use the LR FA algorithm.

\begin{figure}[!t]
\centering
\includegraphics[width=3.5in]{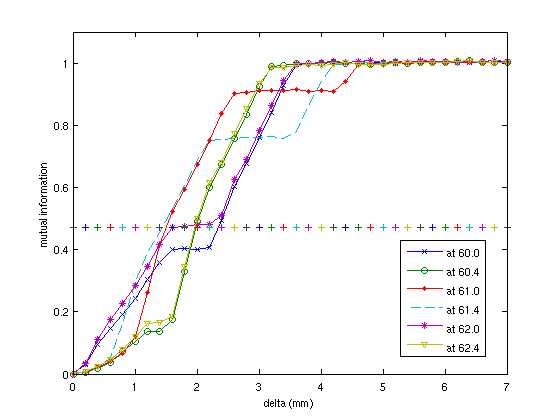}
\caption{Mutual information for the Bidim-26 detector and the FA algorithm with a narrow vertical beam in the sense of the anodes.}
\label{fig_verticalslit_firstt0}
\end{figure}

\begin{figure}[!t]
\centering
\includegraphics[width=3.5in]{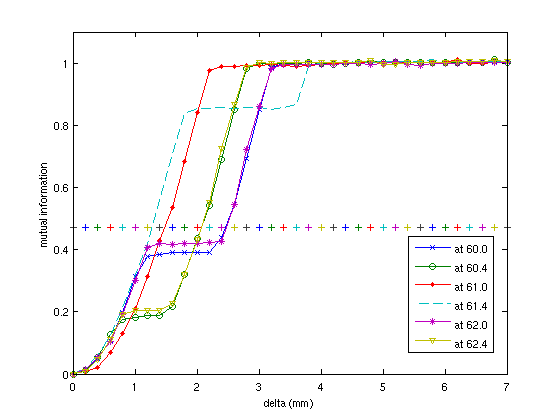}
\caption{Mutual information for the Bidim-26 detector and the max ToT algorithm with a narrow vertical beam in the sense of the anodes.}
\label{fig_verticalslit_maxToT}
\end{figure}

\begin{figure}[!t]
\centering
\includegraphics[width=3.5in]{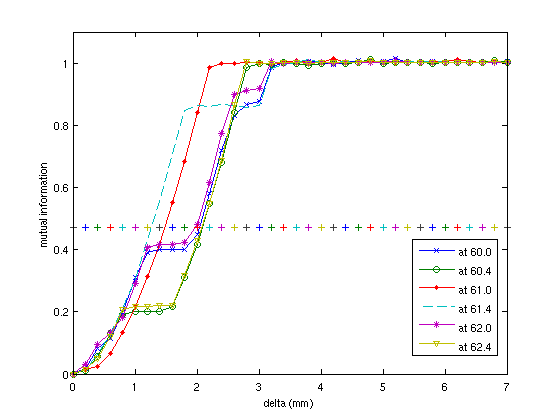}
\caption{Mutual information for the Bidim-26 detector and the CoG ToT algorithm with a narrow vertical beam in the sense of the anodes.}
\label{fig_verticalslit_LRCoGToT}
\end{figure}

The mutual information results for the anode (vertical direction) are shown in figures \ref{fig_verticalslit_firstt0}, \ref{fig_verticalslit_maxToT} and \ref{fig_verticalslit_LRCoGToT}, corresponding respectively to the FA, the max ToT and the CoG ToT algorithms.  The FA algorithm obtains resolutions between 1.4 mm and 2.4 mm ; the max ToT algorithm has a resolution between 1.3 mm and 2.5 mm ; and finally the CoG ToT algorithm between 1.3 mm and 2.1 mm.  It is also interesting to note that the FA algorithm needs up to 4.6 mm to reach in all cases the mutual information of 1 bit, while the max ToT algorithm reaches the 1 bit at 3.8 mm and the CoG ToT algorithm reaches 1 bit at 3.2 mm.

\begin{figure}[!t]
\centering
\includegraphics[width=3.5in]{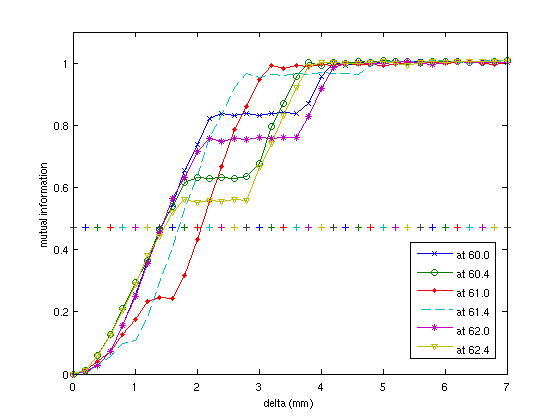}
\caption{Mutual information for the Bidim-26 detector and the FA algorithm with a narrow horizontal beam in the sense of the cathodes.}
\label{fig_horizontalslit_firstt0}
\end{figure}

\begin{figure}[!t]
\centering
\includegraphics[width=3.5in]{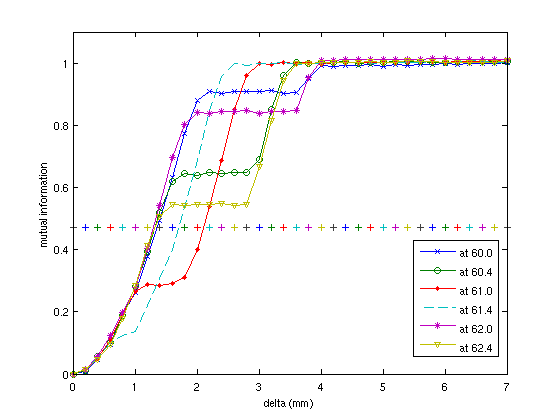}
\caption{Mutual information for the Bidim-26 detector and the max ToT algorithm with a narrow horizontal beam in the sense of the cathodes.}
\label{fig_horizontalslit_maxToT}
\end{figure}

\begin{figure}[!t]
\centering
\includegraphics[width=3.5in]{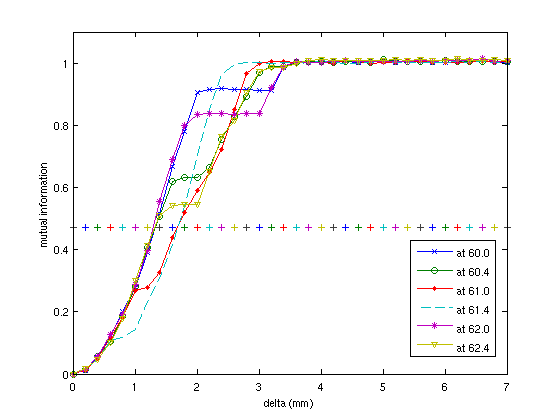}
\caption{Mutual information for the Bidim-26 detector and the CoG ToT algorithm with a narrow horizontal beam in the sense of the cathodes.}
\label{fig_horizontalslit_LRCoGToT}
\end{figure}

For the cathode case, using a narrow horizontal beam, the mutual information curves are shown in figures \ref{fig_horizontalslit_firstt0}, \ref{fig_horizontalslit_maxToT} and \ref{fig_horizontalslit_LRCoGToT}.  We find that the FA algorithm reaches resolutions between 1.4 mm and 2.1 mm, that the max ToT algorithm reaches resolutions between 1.3 mm and 2.1 mm, and that the CoG ToT algorithm reaches resolutions between 1.3 mm and 1.7 mm.  We also observe the superiority of the CoG ToT algorithm over the other two in the distance needed to guarantee 1 bit of mutual information.

The results are better for the cathode where the CoG ToT algorithm can guarantee 1.7 mm than for the anode where 2.1 mm is reached, which corresponds to the fact that the cathode has a higher average multiplicity than the anode, and hence that the discriminator signals contain potentially more information. 

\section{Conclusion}
An information-theoretic definition of resolution has been introduced, which has the advantage of allowing for a meaningful definition of resolution in the case of an imaging system to which the hypothesis of translation-invariance can't be applied on the scale of the resolution.  This definition is based upon considering the image formation system as a lossy communication channel that tries to distinguish between neutron impacts from two different point sources, a certain distance apart.  From the moment that the lossy communication channel reaches a certain threshold of mutual information, we call the needed distance between the two point sources, the local resolution.  This definition goes to the essence of the meaning of image resolution (the capability to distinguish two point sources) and is moreover totally calibration-independent.

This information-theoretical definition is compared to a standard definition of resolution which is based upon the FWHM or the pseudo-FWHM of the PSF.  The resulting equivalent mutual information threshold is 0.47 bits.

The method is then applied to two different neutron detectors with individual readout: a 1D MSGC detector, and a 2D MWPC detector, in order to determine the quality of different online algorithms treating the discriminator signals.  It turns out that the CoG ToT algorithm has the best performance overall, reaching or even slightly improving the best resolution values, and significantly improving the worst resolution values, over the max ToT and FA algorithms.

\end{document}